\documentclass{article}
\usepackage{spconf,amsmath,graphicx,amssymb,latexsym,amsfonts}
\usepackage{verbatim}
\usepackage{bm}

\newtheorem{lemma}{Lemma}

\title{Time-varying Clock Offset Estimation in Two-way Timing Message Exchange in Wireless Sensor Networks Using Factor Graphs$^{\ddag}$}
\name{Aitzaz Ahmad\sthanks{The work of A. Ahmad and E. Serpedin is supported by Qtel.}, Davide Zennaro\sthanks{The work of D. Zennaro is partially supported by an ``A. Gini" fellowship and has been performed while on leave at Texas A\&M University, College Station, TX (USA).\newline{$^{\ddag}$An extended version of this work has been submitted to IEEE Transactions on Information Theory.}}, Erchin Serpedin$^{*}$, Lorenzo Vangelista$^{\dagger}$}
\address{$^{*}$Department
of ECE, Texas A\&M University, Texas,
TX, 77843 USA.\\
$^{\dagger}$Department of Information Engineering, University of Padova, Padova, Italy.}
\begin{document}
%
\maketitle
\begin{abstract}
The problem of clock offset estimation in a two-way timing exchange regime is considered when the likelihood function of the observation time stamps is exponentially distributed. In order to capture the imperfections in node oscillators, which render a time-varying nature to the clock offset, a novel Bayesian approach to the clock offset estimation is proposed using a factor graph representation of the posterior density. Message passing using the max-product algorithm yields a closed form expression for the Bayesian inference problem.
\end{abstract}
\section{Introduction}
\label{sec:intro}

Clock synchronization in wireless sensor networks (WSN) is a critical component in data fusion and duty cycling, and  has gained widespread interest in recent years \cite{akyildiz:survey}. Most of the current methods consider sensor networks exchanging time stamps based on the time at their respective clocks \cite{sadler:survey}. 
In a \emph{two-way} timing exchange process, adjacent nodes aim to achieve pairwise synchronization by communicating their timing information with each other. After a round of $N$ messages, each node tries to estimate its own clock parameters. A representative protocol of this class is the timing-sync protocol for sensor networks (TPSNs) which uses this strategy in two phases to synchronize clocks in a network \cite{ganeriwal:tpsn}.

The clock synchronization problem in a WSN offers a natural statistical signal processing framework \cite{serpedin:survey}. Assuming an exponential delay distribution, several estimators were proposed in \cite{ghaffar:time-out}. It was argued that when the propagation delay $d$ is unknown, the maximum likelihood (ML) estimator for the clock offset $\theta$ is not unique. However, it was shown in \cite{jeske:ML} that the ML estimator of $\theta$ does exist uniquely for the case of unknown $d$. The performance of these estimators was compared with benchmark estimation bounds in \cite{eddie:novel}. A common theme in the aforementioned contributions is that the effect of possible time variations in clock offset, arising from imperfect oscillators, is not incorporated and hence, they entail frequent re-synchronization requirements.

In this work, assuming an exponential distribution for the network delays, a closed form solution to clock offset estimation is presented by considering the clock offset as a random Gauss-Markov process. Bayesian inference is performed using factor graphs and the max-product algorithm.

\section{System Model}
\label{sec:sys}
By assuming that the respective clocks of a sender node $S$ and a receiver node $R$ are related by $C_R(t)=\theta + C_S(t)$ at real time $t$, the two-way timing message exchange model at the $k$th instant can be represented as \cite{ghaffar:time-out} \cite{jeske:ML}
\begin{equation}
U_k=d+\theta+X_k, \quad V_k=d-\theta+Y_k\label{U:V}
\end{equation}
where 
$d$ represents the propagation delay, assumed symmetric in both directions, and $\theta$ is offset of the clock at node $R$ relative to the clock at node $S$. The network delays, $X_k$ and $Y_k$, are the independent exponential random variables. By further defining $\xi\overset{\Delta}{=}d+\theta$ and $\psi\overset{\Delta}{=}d-\theta$, the model in \eqref{U:V} can be written as
\begin{equation}
U_k=\xi+X_k, \quad V_k=\psi+Y_k
\end{equation}
for $k=1,\ldots,N$. The imperfections introduced by environmental conditions in the quartz oscillator in sensor nodes result in a time-varying clock offset between nodes in a WSN. In order to sufficiently capture these temporal variations, the parameters $\xi$ and $\psi$ are assumed to evolve through a Gauss-Markov process given by
\begin{equation*}
\xi_k=\xi_{k-1}+w_k, \quad \psi_k=\psi_{k-1}+v_k \quad \text{for}~k=1,\ldots,N
\end{equation*}
where $w_k$ and $v_k$ are $i.i.d$ such that $w_k,v_k\sim\mathcal{N}(0,\sigma^2)$. The goal is to determine precise estimates of $\xi$ and $\psi$ in the Bayesian framework using observations $\{U_k, V_k\}_{k=1}^{N}$. An estimate of $\theta$ can, in turn, be obtained as
\begin{equation}
\hat\theta = \frac{\hat\xi-\hat\psi}{2}\label{subs:theta}\;.
\end{equation}

\section{A Factor Graph Approach}
\label{sec:factor}
\begin{figure}[t]
\begin{center}
\vspace{0.2in}
\includegraphics[scale=0.5]{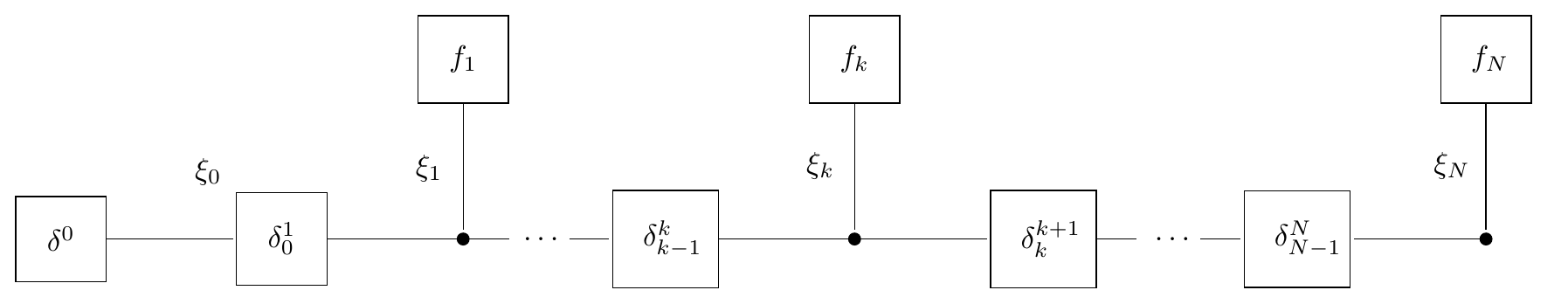}\vspace{0.35in}
\includegraphics[scale=0.5]{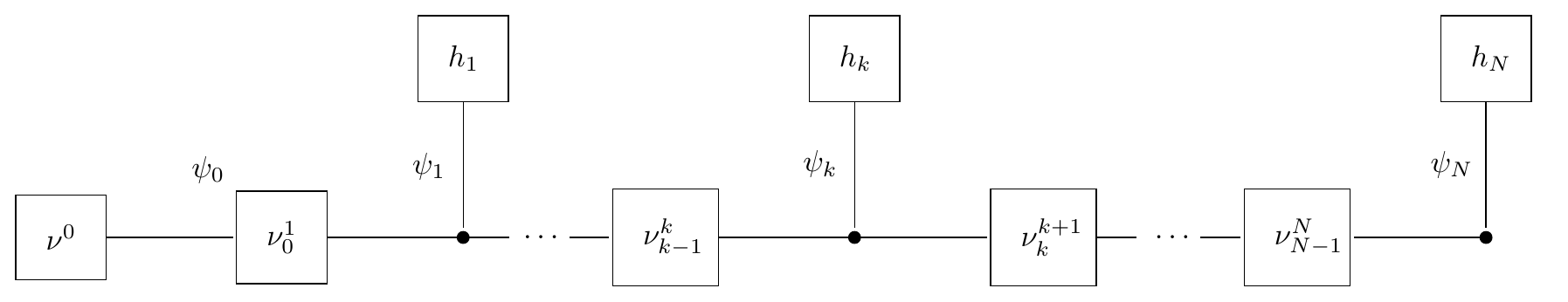}
\caption{Factor graph representation of posterior density \eqref{pdf:pos}}
\label{fig:fact}
\end{center}
\end{figure}

The posterior pdf can be expressed as
\begin{align}
f(\boldsymbol{\xi},\boldsymbol{\psi}|\boldsymbol{U},\boldsymbol{V})&\propto f(\boldsymbol{\xi},\boldsymbol{\psi})f(\boldsymbol{U},\boldsymbol{V}|\boldsymbol{\xi},\boldsymbol{\psi})\nonumber \\
&=f(\xi_0)\prod_{k=1}^{N}f(\xi_k|\xi_{k-1})f(\psi_0)\prod_{k=1}^{N}f(\psi_k|\psi_{k-1})\nonumber \\
&\cdot \prod_{k=1}^{N}f(U_k|\xi_k)f(V_k|\psi_k)\label{pdf:pos}
\end{align}
where uniform priors $f(\xi_0)$ and $f(\psi_0)$ are assumed. Define $\delta_{k-1}^{k}\overset{\Delta}{=}f(\xi_k|\xi_{k-1})\sim\mathcal{N}(\xi_{k-1},\sigma^2)$, $\nu_{k-1}^{k}\overset{\Delta}{=}f(\psi_k|\psi_{k-1})\sim\mathcal{N}(\psi_{k-1},\sigma^2)$, $f_k\overset{\Delta}{=}f(U_k|\xi_k)$, $h_k\overset{\Delta}{=}f(V_k|\psi_k)$, where the likelihood functions are given by
\begin{align}
f(U_k|\xi_k)&=\lambda_{\xi}\exp\left(-\lambda_{\xi}(U_k-\xi_k)\right)\mathbb{I}(U_k-\xi_k)\nonumber\\
f(V_k|\psi_k)&=\lambda_{\psi}\exp\left(-\lambda_{\psi}(V_k-\psi_k)\right)\mathbb{I}(V_k-\psi_k)
\label{pdf:uncons:bayes}
\;.
\end{align}
The factor graph representation of the posterior pdf is shown in Fig. \ref{fig:fact}. The factor graph is cycle-free and inference by message passing is indeed optimal. In addition, the two factor graphs shown in Fig. \ref{fig:fact} have a similar structure and hence, message computations will only be shown for the estimate $\hat\xi_N$. Clearly, similar expressions will apply to $\hat\psi_N$.

The message updates in factor graph using max-product can be computed as follows
\begin{align*}
m_{f_N\rightarrow\xi_N}&=f_N, \quad m_{\xi_N\rightarrow\delta_{N-1}^{N}}=f_N\nonumber \\
m_{\delta_{N-1}^{N}\rightarrow\xi_{N-1}}&\propto \underset{\xi_N}{\max}~\delta_{N-1}^{N}\cdot m_{\xi_N\rightarrow\delta_{N-1}^{N}}\nonumber\\
&=\underset{\xi_N}{\max}~\frac{1}{\sqrt{2\pi\sigma^2}}\exp\left(\frac{-(\xi_N-\xi_{N-1})^2}{2\sigma^2}\right)
\nonumber \\
&\cdot \exp\left(\lambda_{\xi}\xi_N\right) \mathbb{I}(U_N-\xi_N)\label{m:N}
\end{align*}
which can be rearranged as
\begin{equation}
\begin{split}
m_{\delta_{N-1}^{N}\rightarrow\xi_{N-1}}\propto \underset{\xi_N \leq U_N}{\max}~\exp &\big( A_{\xi,N}\xi_N^2+B_{\xi, N}\xi_{N-1}^2+ \\
&C_{\xi, N}\xi_N\xi_{N-1} +D_{\xi, N}\xi_N\big)
\end{split}
\label{m:delta:xi:N-1:max}
\end{equation}
where
\begin{align}
A_{\xi, N}&\overset{\Delta}{=}-\frac{1}{2\sigma^2}, \quad B_{\xi, N}\overset{\Delta}{=}-\frac{1}{2\sigma^2}\nonumber \\
C_{\xi, N}&\overset{\Delta}{=}\frac{1}{\sigma^2}, \quad D_{\xi, N}\overset{\Delta}{=}\lambda_{\xi} \label{const:N}\;.
\end{align}
Let $\bar\xi_N$ be the unconstrained maximizer of the exponent in the objective function above, i.e.,
\begin{equation}
\begin{split}
\bar\xi_N = \arg \underset{\xi_N}{\max}~\big( & A_{\xi, N}\xi_N^2+B_{\xi, N}\xi_{N-1}^2+C_{\xi, N}\xi_N\xi_{N-1}+  \\
& D_{\xi, N}\xi_N \big) \;.\nonumber \\
\end{split}
\end{equation}
This implies that
\begin{equation}
\bar\xi_N=-\frac{C_{\xi, N}\xi_{N-1}+D_{\xi, N}}{2A_{\xi, N}} \label{bar:N}\;.
\end{equation}
If $\bar\xi_N>U_N$, then the estimation problem is solved, since $\hat\xi_N = U_N$. However, if $\bar\xi_N \le U_N$, the solution is $\hat\xi_N = \bar\xi_N$. Therefore, in general, we can write
\begin{equation*}
\hat\xi_N=\min\left(\bar\xi_N,U_N\right)\label{hat:N}\;.
\end{equation*}
Notice that $\bar\xi_N$ depends on $\xi_{N-1}$, which is undetermined at this stage. Hence, we need to further traverse the chain backwards. Assuming that $\bar\xi_{N}\leq U_N$, $\bar\xi_N$ from \eqref{bar:N} can be plugged back in \eqref{m:delta:xi:N-1:max} which after some simplification yields
\begin{equation}
\begin{split}
m_{\delta_{N-1}^{N}\rightarrow\xi_{N-1}}\propto\exp\Bigg\{ &\left(B_{\xi, N}-\frac{C_{\xi, N}^2}{4A_{\xi, N}}\right)\xi^2_{N-1} - \\
& \frac{C_{\xi, N} D_{\xi, N}}{2A_{\xi, N}}\xi_{N-1} \Bigg\} \;.
\end{split}
\label{m:delta:xi:N-1}
\end{equation}
Similarly the message from the factor $\delta_{N-2}^{N-1}$ to the variable node $\xi_{N-2}$ can be expressed as
\begin{align}
m&_{\delta_{N-2}^{N-1}\rightarrow\xi_{N-2}}\propto  \underset{\xi_{N-1} \leq U_{N-1}}{\max}~\delta_{N-2}^{N-1}\cdot m_{\xi_{N-1}\rightarrow\delta_{N-2}^{N-1}}\nonumber\\
=&\underset{\xi_{N-1}}{\max}~\frac{1}{\sqrt{2\pi\sigma^2}}\exp\left(-\frac{(\xi_{N-1}-\xi_{N-2})^2}{2\sigma^2}\right) \nonumber\\
& \cdot\exp\left \{ \left(B_{\xi, N}-\frac{C_{\xi, N}^2}{4A_{\xi, N}}\right)\xi^2_{N-1}-\frac{C_{\xi, N} D_{\xi, N}}{2A_{\xi, N}}\xi_{N-1} \right \}\nonumber\\
&\cdot\exp\left(\lambda_{\xi}\xi_{N-1}\right)\mathbb{I}(U_{N-1}-\xi_{N-1})\nonumber \;.
\end{align}
The message above can be compactly represented as
\begin{align}
m&_{\delta_{N-2}^{N-1} \rightarrow\xi_{N-2}} \propto\nonumber \underset{\xi_{N-1} \leq U_{N-1}} {\max}~\exp(A_{\xi, N-1}\xi_{N-1}^2+ \nonumber \\
&  B_{\xi, N-1}\xi_{N-2}^2+C_{\xi, N-1}\xi_{N-1}\xi_{N-2}+D_{\xi, N-1}\xi_{N-1}) \label{m:delta:xi:N-2:max}
\end{align}
where
\begin{align*}
A_{\xi, N-1}&\overset{\Delta}{=}-\frac{1}{2\sigma^2}+B_{\xi, N}-\frac{C_{\xi, N}^2}{4A_{\xi, N}}, \quad \nonumber \\
B_{\xi, N-1}&\overset{\Delta}{=}-\frac{1}{2\sigma^2}, \quad C_{\xi, N-1}\overset{\Delta}{=}\frac{1}{\sigma^2} \\
D_{\xi, N-1}&\overset{\Delta}{=}\lambda_{\xi}-\frac{C_{\xi, N}D_{\xi, N}}{2A_{\xi, N}}\;.\label{const:N-1}
\end{align*}
Proceeding as before, the unconstrained maximizer $\bar\xi_{N-1}$ of the objective function above is given by
\begin{equation*}
\bar\xi_{N-1}=-\frac{C_{\xi, N-1}\xi_{N-2}+D_{\xi, N-1}}{2A_{\xi, N-1}}\label{bar:N-1}
\end{equation*}
and the solution to the maximization problem \eqref{m:delta:xi:N-2:max} is expressed as
\begin{equation*}
\hat\xi_{N-1}=\min\left(\bar\xi_{N-1},U_{N-1}\right)\;.\label{hat:N-1}
\end{equation*}
Again, $\bar\xi_{N-1}$ depends on $\xi_{N-2}$ and therefore, the solution demands another traversal backwards on the factor graph representation in Fig. \ref{fig:fact}. By plugging $\bar\xi_{N-1}$ back in \eqref{m:delta:xi:N-2:max}, it follows that
\begin{align}
&m_{\delta_{N-2}^{N-1}\rightarrow\xi_{N-2}}\propto\nonumber \\
&\exp\left \{ \left(B_{\xi, N-1}-\frac{C_{\xi, N-1}^2}{4A_{\xi, N-1}}\right)\xi^2_{N-2}-\frac{C_{\xi, N-1} D_{\xi, N-1}}{2A_{\xi, N-1}}\xi_{N-2} \right \}\label{m:delta:xi:N-2}
\end{align}
which has a form similar to \eqref{m:delta:xi:N-1}. It is clear that one can keep traversing back in the graph yielding messages similar to \eqref{m:delta:xi:N-1} and \eqref{m:delta:xi:N-2}. In general, for $i = 1, \ldots, N-1$, we can write
\begin{equation}
\begin{split}
A_{\xi, N-i} &\overset{\Delta}{=} -\frac{1}{2 \sigma^2} + B_{\xi, N-i+1} - \frac{C_{\xi, N-i+1}^2}{4 A_{\xi, N-i+1}} \\
B_{\xi, N-i} & \overset{\Delta}{=} -\frac{1}{2 \sigma^2}, \quad C_{\xi, N-i} \overset{\Delta}{=} \frac{1}{\sigma^2} \\
D_{\xi, N - i} & \overset{\Delta}{=} \lambda_{\xi} - \frac{C_{\xi, N - i + 1} D_{\xi, N - i + 1}}{2 A_{\xi, N - i + 1}}
\end{split}
\label{const:N-i}
\end{equation}
and
\begin{eqnarray}
\bar{\xi}_{N-i} &=& -\frac{C_{\xi, N-i} \xi_{N - i - 1} + D_{\xi, N - i}}{2 A_{\xi, N - i}} \label{bar:N-i} \\
\hat{\xi}_{N-i} &=& \min \left( \bar{\xi}_{N-i}, U_{N - i} \right) \label{hat:N-i}\;.
\end{eqnarray}
Using \eqref{bar:N-i} and \eqref{hat:N-i} with $i = N-1$, it follows that
\begin{equation}
\bar\xi_{1}=-\frac{C_{\xi, 1}\xi_{0}+D_{\xi, 1}}{2A_{\xi, 1}}, \quad \hat\xi_{1}=\min\left(\bar\xi_{1},U_{1}\right) \label{bar:1}\;.
\end{equation}
Similarly, by observing the form of \eqref{m:delta:xi:N-1} and \eqref{m:delta:xi:N-2}, it follows that
\begin{equation}
m_{\delta_{0}^{1}\rightarrow\xi_{0}}\propto \exp\left \{ \left(B_{\xi, 1}-\frac{C_{\xi,1}^2}{4A_{\xi,1}}\right)\xi^2_{0}-\frac{C_{\xi,1} D_{\xi,1}}{2A_{\xi,1}}\xi_{0} \right \}\;.\label{m:delta:xi:0}
\end{equation}
The estimate $\hat\xi_0$ can be obtained by maximizing \eqref{m:delta:xi:0}.
\begin{equation}
\hat\xi_0=\bar\xi_0=\underset{\xi_{0}}{\max}~m_{\delta_{0}^{1}\rightarrow\xi_{0}}
=\frac{C_{\xi,1}D_{\xi, 1}}{4A_{\xi, 1}B_{\xi, 1}-C_{\xi, 1}^2}\;.\label{hat:0}
\end{equation}
The estimate in \eqref{hat:0} can now be substituted in \eqref{bar:1} to yield $\bar\xi_1$, which can then be used to solve for $\hat\xi_1$. Clearly, this chain of calculations can be continued using recursions \eqref{bar:N-i} and \eqref{hat:N-i}.\\
Define
\begin{equation}
g_{\xi,k}(x)\overset{\Delta}{=}-\frac{C_{\xi,k}x+D_{\xi,k}}{2A_{\xi,k}}\;.\label{g:def}
\end{equation}
\begin{lemma}
For real \label{lem:1}numbers $a$ and $b$, the function $g_{\xi,k}(.)$ defined in \eqref{g:def} satisfies
\begin{equation*}
g_{\xi,k}\left(\min(a, b)\right)=\min\left(g_{\xi,k}(a), g_{\xi,k}(b)\right)\;.
\end{equation*}
\end{lemma}
Proof: The constants $A_{\xi,k}$, $C_{\xi,k}$ and $D_{\xi,k}$ are defined in \eqref{const:N} and \eqref{const:N-i}. The proof follows by noting that $\frac{-C_{\xi,k}}{2A_{\xi,k}}>0$ which implies that $g_{\xi,k}(.)$ is a monotonically increasing function.\\

\noindent Using the notation $g_{\xi,k}(.)$, it follows that
\begin{align*}
\bar\xi_1&=g_{\xi,1}\left(\hat\xi_0\right), \quad \hat\xi_1=\min\left(U_1,g_{\xi,1}\left(\hat\xi_0\right)\right)\nonumber \\
\bar\xi_2&=g_{\xi,2}\left(\hat\xi_1\right), \quad \hat\xi_2=\min\left(U_2,g_{\xi,2}\left(\hat\xi_1\right)\right)\nonumber
\end{align*}
where
\begin{align}
g_{\xi,2}\left(\hat\xi_1\right)&=g_{\xi,2}\left(\min\left(U_1,g_{\xi,1}\left(\hat\xi_0\right)\right)\right)\nonumber \\
&=\min\left(g_{\xi,2}\left(U_1\right),g_{\xi,2}\left(g_{\xi,1}\left(\hat\xi_0\right)\right)\right)\label{g:nest}
\end{align}
where \eqref{g:nest} follows from Lemma \ref{lem:1}. The estimate $\hat\xi_2$ can be expressed as
\begin{align*}
\hat\xi_2&=\min\left(U_2,\min\left(g_{\xi,2}\left(U_1\right),g_{\xi,2}\left(g_{\xi,1}\left(\hat\xi_0\right)\right)
\right)\right)
\nonumber\\
&=\min\left(U_2,g_{\xi,2}\left(U_1\right),g_{\xi,2}\left(g_{\xi,1}\left(\hat\xi_0\right)\right)\right)\;.
\end{align*}
Hence, one can keep estimating $\hat\xi_k$ at each stage using this strategy. Note that the estimator only depends on functions of data and can be readily evaluated. For $m \geq k$, define
\begin{equation}
G_{\xi, k}^m(.)\overset{\Delta}{=}g_{\xi,m}\left(g_{\xi,m-1}\ldots g_{\xi,k}\left(.\right)\right)\;.\label{G:def}
\end{equation}
The estimate $\hat\xi_N$ can, therefore, be compactly represented as
\begin{equation}
\hat\xi_N=\min\left(U_N,G_{\xi,N}^N\left(U_{N-1}\right),\dots,G_{\xi,2}^N\left(U_1\right),
G_{\xi,1}^N\left(\hat\xi_0\right)\right) \label{xi:1}\;.
\end{equation}
By a similar reasoning, the estimate $\hat\psi_N$ can be analogously expressed as
\begin{equation*}
\hat\psi_N=\min\left(V_N,G_{\psi,N}^N\left(V_{N-1}\right),\dots,G_{\xi,2}^N\left(V_1\right),
G_{\xi,1}^N\left(\hat\psi_0\right)\right)
\end{equation*}
and the factor graph based clock offset estimate (FGE) $\hat\theta_N$ is given by
\begin{equation}
\hat\theta_N=\frac{\hat\xi_N-\hat\psi_N}{2}\;.\label{theta:fg}
\end{equation}
It only remains to calculate the functions of data $G(.)$ in the expressions for $\hat\xi_N$ and $\hat\psi_N$ to determine the FGE estimate $\hat\theta_N$. With the constants defined in \eqref{const:N}, it follows that
\begin{equation*}
G_{\xi,N}^N(U_{N-1})=-\frac{C_{\xi,N}U_{N-1}+D_{\xi,N}}{2A_{\xi,N}}=U_{N-1}+\lambda_{\xi}\sigma^2\;.
\end{equation*}
Similarly it can be shown that
\begin{equation*}
G_{\xi,N-1}^N(U_{N-2})=U_{N-2}+2\lambda_{\xi}\sigma^2
\end{equation*}
and so on. Using the constants defined in \eqref{const:N-i} for $i=N-1$, it can be shown that $\hat\xi_0=\frac{C_{\xi,1}D_{\xi,1}}{4A_{\xi,1}B_{\xi,1}-C_{\xi,1}^2}=+\infty$. This implies that $G_{\xi,1}^N(\hat\xi_0)=+\infty$. Plugging this in \eqref{xi:1} yields
\begin{equation*}
\hat\xi_N=\min(U_N, U_{N-1}+\lambda_{\xi}\sigma^2,
\ldots,U_1+(N-1)\lambda_{\xi}\sigma^2)\;.\label{map:E}
\end{equation*}
Similarly, the estimate $\hat\psi_N$ is given by
\begin{equation*}
\hat\psi_N=\min(V_N, V_{N-1}+\lambda_{\psi}\sigma^2,
\ldots,V_1+(N-1)\lambda_{\psi}\sigma^2)\;\label{map:E:psi}
\end{equation*}
and the estimate $\hat\theta_N$ can be obtained using \eqref{theta:fg} as
\begin{align}
\hat\theta_N=\frac{1}{2} \min(U_N, U_{N-1}&+\lambda_{\xi}\sigma^2,U_{N-2}+2\lambda_{\xi}\sigma^2,\nonumber \\
&\ldots,U_1+(N-1)\lambda_{\xi}\sigma^2)-\nonumber\\\frac{1}{2}\min(V_N, V_{N-1}&+\lambda_{\psi}\sigma^2,V_{N-2}+2\lambda_{\psi}\sigma^2,\nonumber \\
&\ldots,V_1+(N-1)\lambda_{\psi}\sigma^2)\;.\label{example:exp:theta:N}
\end{align}
As the Gauss-Markov system noise $\sigma^2 \rightarrow 0$, \eqref{example:exp:theta:N} yields
\begin{equation}
\hat\theta_N\rightarrow \hat\theta_{\textrm{ML}}=\frac{\min\left(U_N, \ldots, U_1\right)-\min\left(V_N, \ldots, V_1\right)}{2}\label{ml}
\end{equation}
which is the ML estimator proposed in \cite{jeske:ML}.
\begin{figure}
\centering
\includegraphics[scale=0.35]{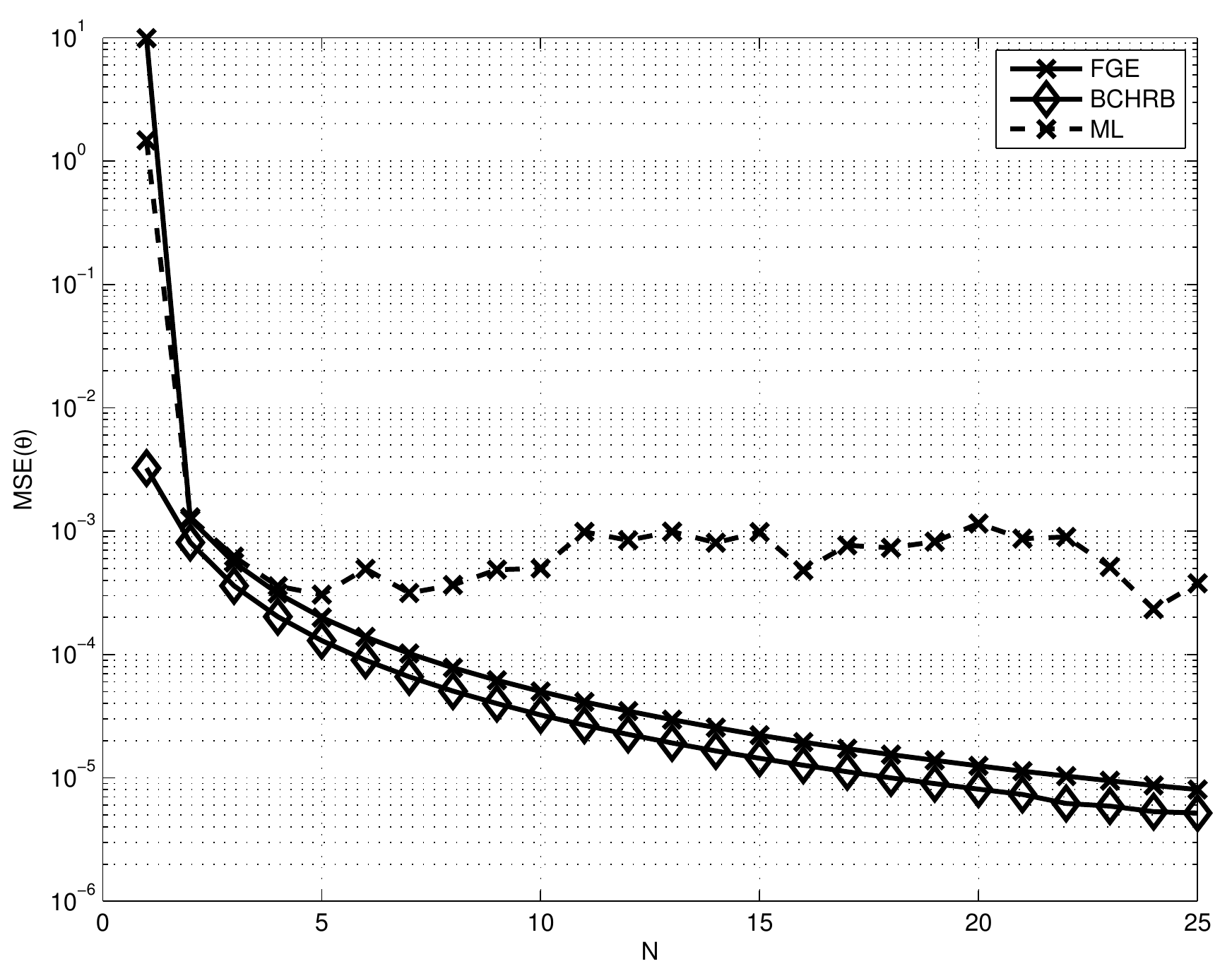}
\caption{Comparison of MSE of $\hat\theta_N$ and $\hat\theta_{\textrm{ML}}$.}
\label{Bayes_theta_eps}
\end{figure}

\section{Simulation Results}
\begin{figure}[t]
\centering
\includegraphics[scale=0.35]{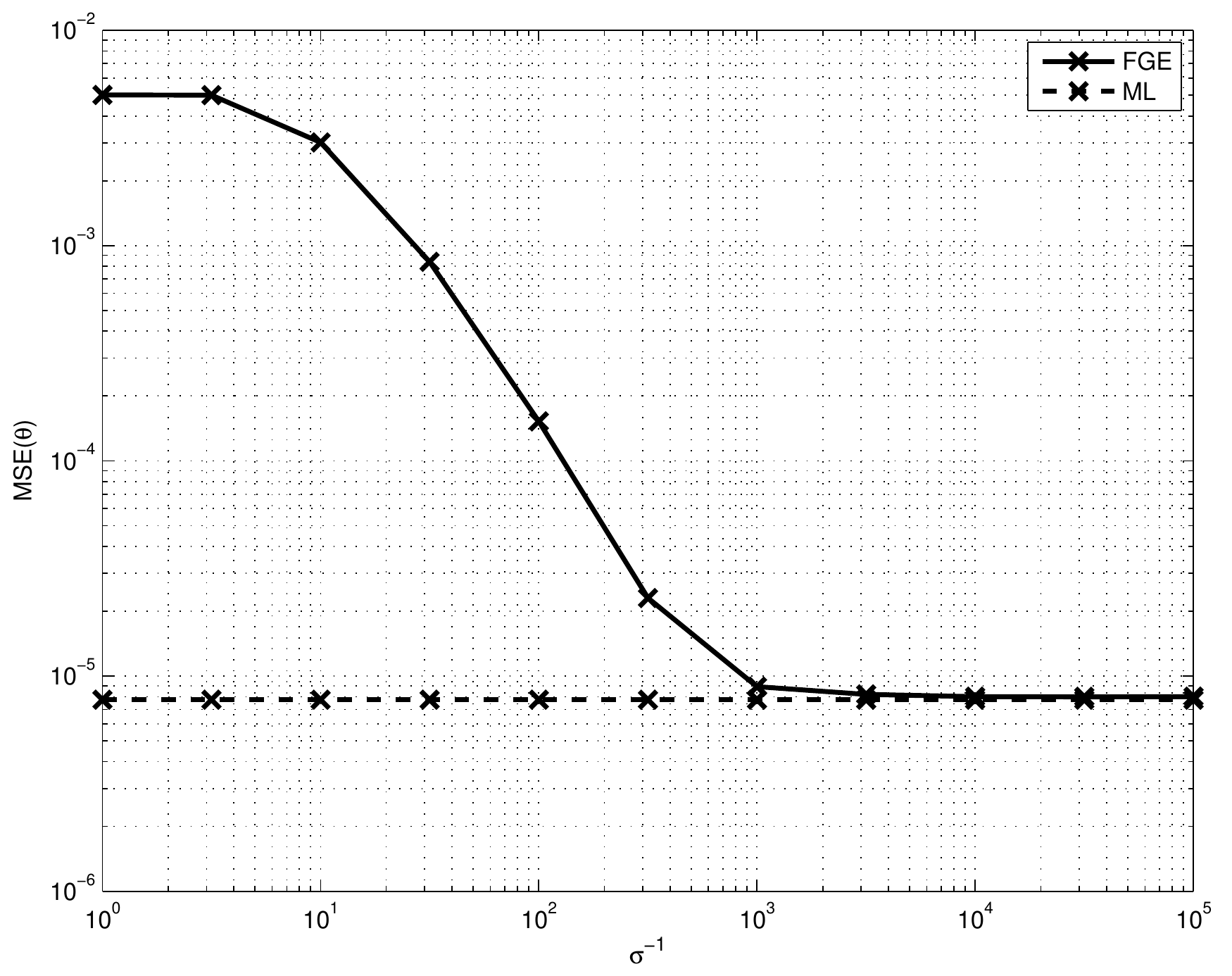}
\caption{MSE in estimation of $\theta_N$ vs $\sigma$.}
\label{Bayes_theta_vs_sigma_GM}
\end{figure}

With $\lambda_{\xi}=\lambda_{\psi}=10$ and $\sigma=10^{-2}$, Fig. \ref{Bayes_theta_eps} shows the MSE performance of $\hat\theta_N$ and $\hat\theta_{\textrm{ML}}$, compared with the Bayesian Chapman-Robbins bound (BCHRB). It is clear that $\hat\theta_N$ exhibits a better performance than $\hat\theta_{\textrm{ML}}$ by incorporating the effects of time variations in clock offset. As the variance of the Gauss-Markov model accumulates with the addition of more samples, the MSE of $\hat\theta_{\textrm{ML}}$ gets worse. Fig. \ref{Bayes_theta_vs_sigma_GM} depicts the MSE of $\hat\theta_N$ in \eqref{example:exp:theta:N} with $N = 25$. The horizontal line represents the MSE of the ML estimator \eqref{ml}. It can be observed that the MSE obtained by using the FGE for estimating $\theta$ approaches the MSE of the ML as $\sigma < 10^{-3}$.

\section{Conclusion}
The estimation of a possibly time-varying clock offset is studied using factor graphs. A closed form solution to the clock offset estimation problem is presented using a novel message passing strategy based on the max-product algorithm. This estimator shows a performance superior to the ML estimator proposed in \cite{jeske:ML} by capturing the effects of time variations in the clock offset efficiently.

\end{document}